\documentclass[%
 preprint,
 amsmath,amssymb,
 aps,
]{revtex4-2}

\usepackage{graphicx}% Include figure files
\usepackage{dcolumn}% Align table columns on decimal point
\usepackage{bm}% bold math
\usepackage{hyperref}% add hypertext capabilities
\usepackage{siunitx}
\begin{document}

\preprint{APS/123-QED}

\title{Electron-beam induced methane decomposition for in-situ carbon doping of hexagonal boron nitride}% Force line breaks with\\

\author{Barbara Maria Mayer$^{1,2}$}
\author{Manuel Längle$^{1,3}$}
\author{Umair Javed$^{1,4}$}
\author{Toma Susi$^{1}$}
\author{E. Harriet \AA hlgren$^{2}$}
\author{Jani Kotakoski$^{1}$}

\affiliation{%
 $^{1}$University of Vienna, Faculty of Physics, Boltzmanngasse 5, 1090 Vienna, Austria}

\affiliation{$^{2}$ Uppsala University, Box 516, Uppsala, SE-751 20, Sweden}

\affiliation{$^{3}$ Univ. Paris-Saclay, CNRS, Laboratoire de Physique des Solides,
91405, Orsay, France}

\affiliation{$^{4}$ University of Vienna, Vienna Doctoral School in Physics,
Boltzmanngasse 5, 1090 Vienna, Austria}

\date{\today}% It is always \today, today,

\begin{abstract}

Controlling the spatial incorporation of carbon into hexagonal boron nitride (hBN) is essential for engineering optically active defects, yet existing approaches lack nanoscale precision and control over the carbon supply. Here, we demonstrate a method for carbon doping of hBN using electron-beam irradiation in a low-pressure methane atmosphere, where the beam simultaneously generates vacancies and decomposes methane into individual carbon and hydrogen atoms. Using annular dark-field scanning transmission electron microscopy, we show that increasing the methane partial pressure suppresses pore growth and drives the formation of triangular boron-terminated pores through preferential hydrogen etching of nitrogen. Time-resolved electron energy-loss spectroscopy (EELS) mapping reveals progressive carbon incorporation into the lattice, accompanied by boron and nitrogen depletion. Carbon clustering occurs predominantly within the irradiated area: 84$\pm$7\% of carbon-rich regions are confined to the area exposed to the electron beam, while some carbon atoms are also found to diffuse up to an average distance of 4.7$\pm$0.5~nm beyond it. The incorporated carbon atoms arrange in a hexagonal pattern within the lattice, forming patches that do not exceed $\sim$1~nm in size. Analysis of the EELS fine structure indicates modifications to the local electronic environment within these regions, with implications for the optical properties of the resulting carbon-related defects.

%26.April:Point defects, such as vacancies or substitutional impurity atoms, in monolayer hexagonal boron nitride (hBN) can serve as stable single-photon emitters at room temperature. Among these, carbon centers are particularly significant for visible-light photon emission, but methods for controlling carbon incorporation remain limited. Here, we demonstrate that electron-beam irradiation in a low-pressure methane atmosphere enables localized incorporation of carbon atoms into hBN. Using annular dark-field scanning transmission electron microscopy, electron ptychography, and electron energy-loss spectroscopy, we observe how carbon integrates into the lattice and influences defect formation. We find that increasing the carbon partial pressure suppresses electron-irradiation-induced pore growth and introduces a new type of defect: an electron-beam-induced triangular pore that is not nitrogen-terminated, unlike those created under a low-pressure oxygen atmosphere. Simultaneously, carbon atoms gradually replace boron and nitrogen, creating carbon-containing lattice regions. When these regions reach approximately two nanometers in diameter, we observe changes in the local electronic structure. Finally, we demonstrate that carbon incorporation is highly localized at the beam position, enabling nanometer-precision C doping of hBN. 
\end{abstract}

\maketitle

\section*{Introduction}

Two-dimensional (2D) hexagonal boron nitride (hBN) is a one-atom-thick material composed of boron and nitrogen atoms in a honeycomb lattice. The alternating arrangement of two atomic species with different electronegativities leads to a large electronic band gap that can be exploited to create localized states by introducing point defects, such as vacancies or substitutional impurity atoms. These defects can act as single-photon emitters (SPEs) when the material is optically excited, and are particularly attractive because they can exhibit spectrally narrow emission lines~\cite{Akbari2022Lifetime-LimitedModulation, Dietrich2020Solid-stateTemperature}, ranging from ultraviolet to near-infrared, and demonstrate excellent photostability~\cite{Li2023ProlongedEmitters} even at room temperature~\cite{Grosso2017TunableNitride}. As a result, point defects in hBN are promising for applications in quantum communication~\cite{Hanson2008CoherentSemiconductors} and computing~\cite{OBrien2007OpticalComputing}, as well as sensing~\cite {Schirhagl2014Nitrogen-vacancyBiology}.

While linking specific photon-emission signatures to particular defective atomic structures remains debated, in part because of challenges in their microscopic characterization~\cite{Lamprecht2026SingleError}, a growing consensus suggests that, apart from the negatively charged boron vacancy~\cite{Ivady2020AbNitride}, most single-photon emission in hBN originates from carbon-related defects~\cite{Vogl2026DefectsPerspective, Bourrellier2016BrightH-BN, Zhigulin2023PhotophysicsNitride, Tran2017RobustNitride, Tawfik2017First-principlesDefects, Mendelson2021IdentifyingNitride}. Consequently, methods for deliberately introducing carbon atoms into the hBN lattice are of significant interest. One approach involves incorporating carbon during material synthesis~\cite{Mendelson2021IdentifyingNitride, Williams2025QuantumHBN, Liu2022RationalNitride}, enabling large-scale production of carbon-doped hBN with a relatively uniform carbon distribution~\cite{Tang2025Structured-DefectEmitters}. However, such bottom-up techniques offer limited control over the precise placement of individual carbon atoms. This limitation has motivated the development of post-synthesis doping methods that target specific regions after crystal formation. Techniques such as ion implantation~\cite{Mendelson2021IdentifyingNitride, Zhong2024Carbon-RelatedPolarization}, post-growth carbon diffusion into vacancies~\cite{Ngamprapawat2023FromDevices}, and focused-ion-beam patterning~\cite{Wu2025Site-ControlledNitride} can achieve site-specific doping, but they often induce lattice damage, require high processing temperatures, or involve complex fabrication steps. To address these issues, electron-beam-assisted carbon doping has emerged as an alternative for introducing carbon into hBN~\cite{Wei2011Electron-beam-inducedNanotubes, Wei2010Post-synthesisIrradiation, Park2021AtomicallyGuide}. Notably, Park et al.~\cite{Park2021AtomicallyGuide} demonstrated the atomically precise insertion of carbon atoms from surface contamination into pre-existing point vacancies in hBN using a focused electron beam. While this work established atomic-scale control at individual defect sites, it relies on an uncontrolled carbon supply and on pre-existing defect sites, which limit control over the spatial extent and distribution of carbon-doped regions---an essential requirement for future applications of C-doped hBN.

To address these limitations, we introduce a method that simultaneously creates lattice vacancies and supplies carbon from a controlled source. Methane (CH$_4$) gas is introduced directly into the column of a scanning transmission electron microscopy (STEM) instrument, where the focused electron beam dissociates the molecules into atomic carbon and hydrogen. The same beam that drives dissociation also ejects boron and nitrogen atoms from the lattice~\cite{Bui2023CreationIrradiation}, creating vacancies and small pores~\cite{Kotakoski2010ElectronMonolayers, JavedInfluenceIrradiation} into which carbon can incorporate. This pathway allows us to control both the vacancy generation rate and the carbon supply without relying on pre-existing defect sites or an uncontrolled reservoir of contamination. We characterize this process using three complementary experimental approaches. First, we quantify how methane partial pressure alters pore growth rates and morphology using time-resolved medium-angle annular dark-field (MAADF) image stacks, revealing a transition from uncontrolled pore growth due to oxygen etching of boron~\cite{JavedInfluenceIrradiation} to saturation and boron-terminated triangular pore formation driven by preferential nitrogen etching by hydrogen. Second, time-resolved EELS maps tracking the spatial and temporal evolution of boron, carbon, and nitrogen concentrations show that carbon progressively incorporates into the lattice as nanoscale patches, predominantly within the electron-beam-irradiated region. Third, high-resolution imaging after extended irradiation reveals the atomic arrangement of carbon in these patches, showing a hexagonal substitutional geometry with a characteristic size of $\sim$1~nm. Taken together, our findings propose a route to spatially controlled, nanoscale carbon doping of hBN, with promising prospects for the deterministic engineering of carbon-related single-photon emitters.

\section*{Results and Discussion}

\begin{figure*}[ht]
	\centering
	\includegraphics[width=\textwidth]{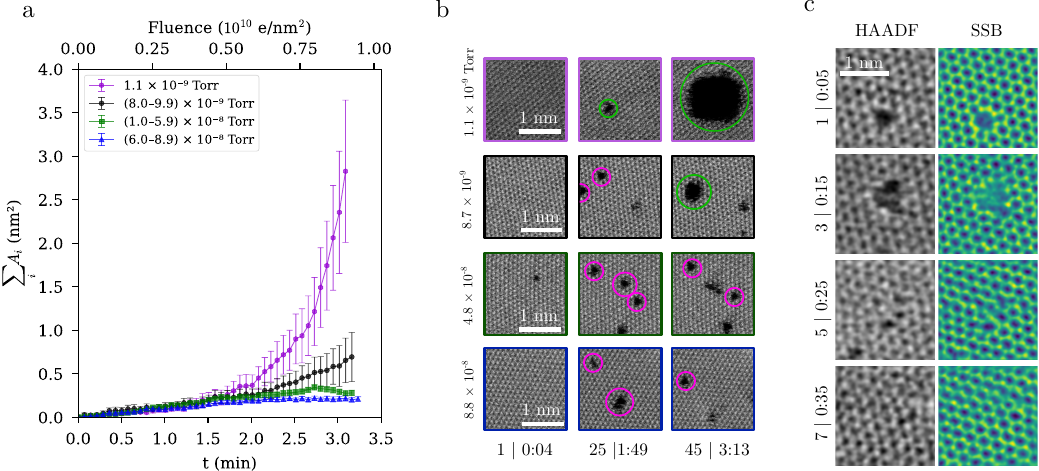}
	\caption{Pore growth in an CH$_4$ atmosphere. (a) Total pore area within a $4\times4$ nm$^2$ field of view as a function of time. Values are averaged over MAADF image stacks recorded at each pressure regime, with error bars indicating the standard error of the mean. The lowest pressure corresponds to no added methane in the column. (b) Selected frames from time-resolved MAADF image stacks recorded at 1.1$\times$10$^{-9}$~Torr (purple frames), 8.7$\times$10$^{-9}$ Torr (black frames), 4.8$\times$10$^{-8}$ Torr (green frames), and 8.8$\times$10$^{-8}$ Torr (blue frames), shown for the first recorded frame (0:04 min), the 25th frame (1:49 min), and the 45th frame (3:13 min). The green-circled defect corresponds to a pore with mixed boron and nitrogen termination, while the purple circles highlight pores with a triangular character and boron termination. (c) Selected frames from a time series showing concurrent filtered HAADF images (left column) and single-sideband (SSB) phase reconstructions (right column, false-colored with the viridis colormap) of pores at a methane partial pressure of 5.5$\times 10^{-8}$~Torr, illustrating pore growth followed by filling.
    } \label{fig_growth}
 %At pressures of 8.0--9.9 $\times$ 10$^{-9}$ Torr, the total pore area in a 4 $\times$ 4 nm$^2$ frame grows quasi-quadratically, while in the pressure range  1.0--5.9 $\times$ 10$^{-8}$ Torr, the total pore area grows less, and sometimes even decreases. In the range 1.0--5.9 $\times$ 10$^{-8}$ Torr, the  total pore area reaches a saturation point after $\sim$ 1.6.% } \label{fig_growth}
\end{figure*}

Samples were prepared by electrochemical delamination of chemical vapor deposition (CVD)–grown hBN from the copper substrate, followed by transfer onto Quantifoil$^{\mathrm{TM}}$ Au TEM grids with 1.2 \textmu m holes, as described in Ref.~\cite{Irschik2026AtomicallyVacuum}. Before insertion into the microscope, the samples were baked in vacuum at ca. 150$^\circ$C for approximately 12 h.

For the experiments, the Vienna Nion UltraSTEM 100 was operated at 60 kV. While this microscope has a base pressure of $\sim$1.5$\times$10$^{-10}$~Torr in the objective area~\cite{Leuthner2019ScanningAtmospheres}, at the time of these experiments, one of the bellows at the objective area had a minor leak, leading to a base pressure roughly an order of magnitude higher than this. Under these conditions, we expect the major molecular species in the residual vacuum to be water. Nevertheless, a leak valve mounted on a port near the objective area allows controlling the local atmosphere around the sample~\cite{Ahlgren2022Atomic-ScaleMoTe2, Leuthner2019ScanningAtmospheres, JavedInfluenceIrradiation} without disrupting imaging up to pressures around $10^{-6}$~Torr. To introduce gas in a controlled manner, the pressure in the objective area is monitored by a gauge; however, the actual pressure at the sample is estimated to be ten times higher than the gauge reading~\cite{Leuthner2019ScanningAtmospheres}. All pressures stated here are from the objective-area gauge.

%23.AprilIn the experiments,the Vienna Nion UltraSTEM 100 was used, operated at 60kV. One modification involves adding a leak valve mounted on a port near the sample area, enabling controlled gas introduction directly into the microscope's objective region~\cite{Ahlgren2022Atomic-ScaleMoTe2, Leuthner2019ScanningAtmospheres, JavedInfluenceIrradiation}. When the leak valve is closed, the objective area maintains a base pressure of $\sim$2 $\times$ 10$^{-10}$ mbar~\cite{Leuthner2019ScanningAtmospheres}. Opening the valve introduces gas into the objective region. The pressure in this area is monitored by a gauge; however, the actual pressure at the sample is estimated to be ten times higher than the gauge reading~\cite{Leuthner2019ScanningAtmospheres}. All pressures stated here are taken from the objective area gauge.

%Because it was recently discovered that oxygen in the sample chamber promotes faster pore growth and the formation of triangular, nitrogen-terminated pores~\cite{JavedInfluenceIrradiation}, we began our experiments by examining how methane influences pore growth. 

In the first set of experiments, we examined how methane affects pore growth. This study was motivated by recent findings that oxygen in the objective area accelerates pore growth and leads to the formation of the well-known triangular, nitrogen-terminated pores \cite{JavedInfluenceIrradiation, Kotakoski2010ElectronMonolayers}. In contrast, the role of hydrocarbons---such as methane, a common residual gas in vacuum chambers---has not yet been explored. Here, methane was introduced into the microscope objective area until partial pressures between (8.0 to 8.9)$\times$10$^{-8}$ Torr were reached. Once the pressure stabilized, time-resolved medium-angle annular dark-field (MAADF) image stacks were recorded over a 4$\times$4 nm$^2$ field of view (FOV) with a dwell time of 16~\textmu s and a size of 512$\times$512 pixels. Each image was then analyzed semi-automatically using intensity thresholding to measure the total pore area over time.

As shown in Fig.~\ref{fig_growth}(a), the pore-growth rate depends strongly on the methane partial pressure within the column. Before methane leakage, the pressure in the objective area was 1.1$\times$10$^{-9}$ Torr, and pore growth exhibited a quadratic trend. This behavior can be attributed to chemical etching by residual oxygen (either from water molecules or from molecular oxygen) in the column~\cite{JavedInfluenceIrradiation}. When methane is introduced at pressures of (8.0 to 9.9)$\times$10$^{-9}$ Torr, similar quadratic behavior remains, but the pore-growth rate decreases. However, a substantial qualitative change in the growth rate occurs when the methane partial pressure is increased to (1.0 to 5.9)$\times$10$^{-8}$ Torr. Now, the pores tend to stabilize over time and eventually reach saturation, and, notably, in some cases, decrease in size, indicating partial pore filling. Further increases in methane partial pressure lead to even stronger suppression of pore growth. The pore-growth rate in this regime is comparable to the behavior reported for graphene under electron irradiation in ultra-high vacuum~\cite{Leuthner2021ChemistryMicroscope, Kotakoski2012StabilityEffects}.

In addition to affecting the pore-growth rate, variations in methane partial pressure within the column also cause distinct pore morphologies. At a pressure of 1.1$\times$10$^{-9}$ Torr, pores with boron- and nitrogen-terminated edges and a slight triangular character develop over time, as shown in the purple-enframed image sequence in Fig.~\ref{fig_growth}(b). The slight triangular character can be explained by preferential etching of boron by residual oxygen~\cite{JavedInfluenceIrradiation}. When methane is introduced into the objective area, the pore morphology changes. At a pressure of 8.7$\times$10$^{-9}$ Torr, shown in the black-enframed image sequence in Fig.~\ref{fig_growth}(b), triangular and metastable boron-terminated pores appear; however, they evolve into rounder pores with a mixed termination. Upon further increasing the methane partial pressure, relatively stable triangular boron-terminated pores form, as shown in the sequences framed in blue and green in Fig.~\ref{fig_growth}(b), recorded at 4.8$\times$10$^{-8}$ and 8.8$\times$10$^{-8}$ Torr, respectively, though structural reconstruction and filling of pores is also observed, as shown in Fig.~\ref{fig_growth}(c). In this image sequence, we observe a triangular pore corresponding to a tetravacancy forming first, growing somewhat over the next couple of images, and starting to fill, reaching a closed structure by frame 7 after only 35~s. The single-sideband ptychography images (shown in false color) allow the arrangement of atoms to be deduced in each frame, whereas the also shown HAADF images are less clear.

%MAADF images indicate that pores formed in a methane environment terminate at atoms belonging to the boron sublattice. However, because ADF imaging at the high frame rates required to capture pore dynamics is inherently noisy, it was not possible to reliably distinguish between B and C, thereby necessitating electron ptychography. Although single-sideband ptychography is more dose-efficient and can clearly resolve the B and N sublattices, as shown in Fig.~\ref{fig_triangle}, the signal remains limited for monolayer low-$Z$ hBN at the shortest available pixel dwell time of 10 $\mu$s. Furthermore, its nonlinear contrast-formation mechanism~\cite{hofer_reliable_2023} makes reliable elemental identification of undercoordinated edge atoms particularly challenging

% 20.May:We tried to resolve this ambiguity by using electron ptychography, given its distinct contrast-forming mechanism compared with ADF imaging. Although single-sideband ptychography is more dose-efficient and can reliably distinguish the B and N sublattices, as shown in Fig.~\ref{fig_triangle}, at the fastest available pixel dwell time of 10~\textmu s the signal for a low-$Z$ hBN is still limited, and its nonlinear contrast-formation mechanism~\cite{hofer_reliable_2023} makes reliable elemental identification at undercoordinated edge atoms likewise challenging.

\begin{figure}
	\centering
	\includegraphics[width=5.5cm]{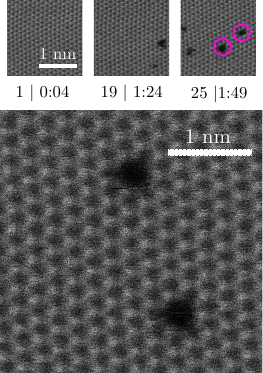}
	\caption{MAADF images acquired at a hydrogen partial pressure of 1.1$\times$10$^{-7}$ Torr. The upper row shows the first (0:04 min), 19th (1:24 min), and 25th (1:49 min) recorded frames of an image series. The purple-encircled defects indicate triangular boron-terminated pores, two of which are clearly visible in the high-resolution MAADF image at the bottom.\label{fig_triangle}}
\end{figure}

While it is natural to expect the filling of the pores to be related to carbon released by the dissociation of the methane molecules under the electron beam, it is much less clear what the cause is for the formation of triangular pores that terminate at the boron sublattice. Indeed, from either the ADF or SSB images, at the fast imaging parameters required to observe pore growth, it was difficult to reliably determine whether the atomic species at the edge is boron or carbon. Therefore, to better understand this, we carried out control experiments under a hydrogen atmosphere at 1.1$\times$10$^{-7}$~Torr. As shown in Fig.~\ref{fig_triangle}, under continuous electron irradiation, metastable triangular pores (mostly as a tetravacancy) with the same edge termination as those observed in a methane environment are present. The increased stability of these structures under a hydrogen atmosphere also allows recording ADF images with much higher signal-to-noise, unambiguously showing that the edge atoms are boron. This suggests that hydrogen preferentially etches nitrogen and that the triangular pores observed under methane are terminated by boron rather than nitrogen.

\begin{figure*}[t]
	\centering
	\includegraphics[width=\textwidth]{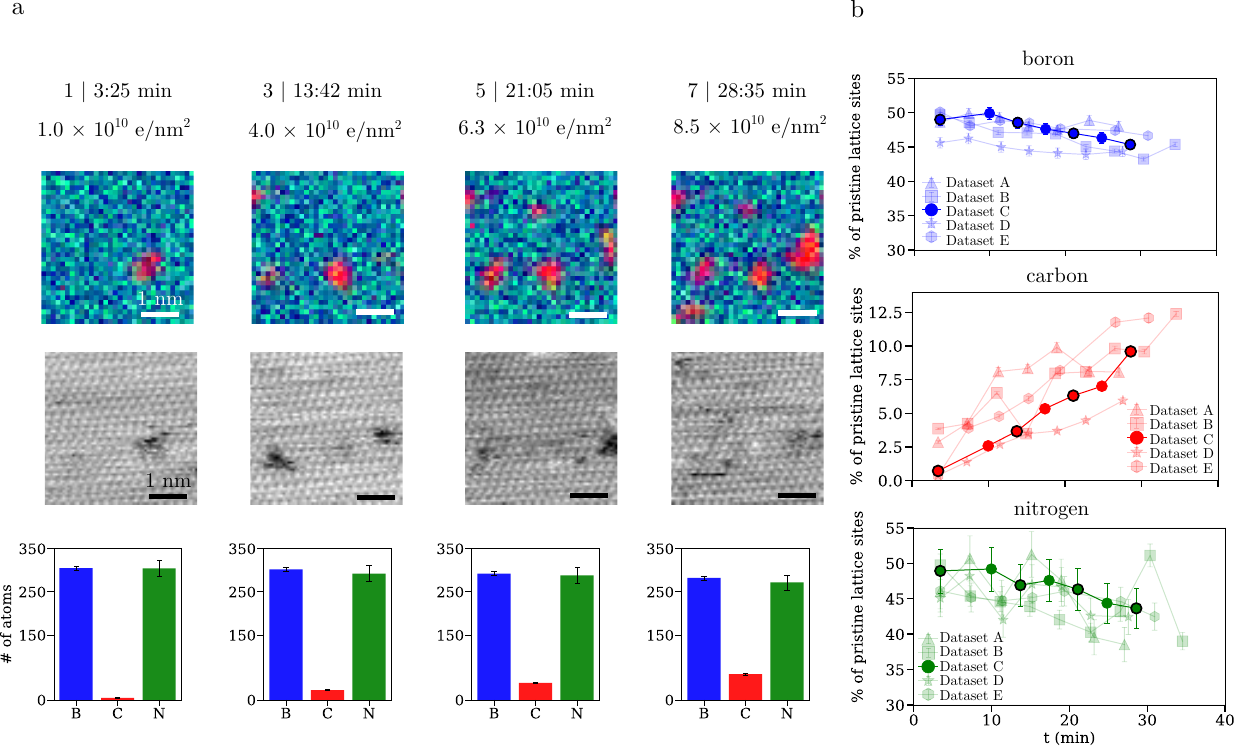}
	\caption{Temporal evolution of carbon incorporation and concurrent boron and nitrogen depletion in hBN under electron-beam irradiation in a methane atmosphere at (1.0 to 1.9)$\times 10^{-7}$~Torr. (a) The first row shows elemental maps derived from the integrated B--, C--, and N--$K$ edge signals, shown in blue, red, and green, respectively. The second row presents the corresponding HAADF images recorded simultaneously. Scan artifacts are due to sample drift during the slow acquisition. The third row displays the numbers of B, C, and N atoms extracted from each frame, illustrating the progressive incorporation of carbon and the reduction of boron and nitrogen over time. (b) Relative areal fractions of boron, carbon, and nitrogen over time for five different datasets. A pristine hBN lattice corresponds to 50$\%$ boron and 50$\%$ nitrogen. The dataset shown in panel (a) is highlighted with opaque markers, while the remaining datasets are shown with semi-transparent symbols. Black-circled data points indicate the frames corresponding to those displayed in (a).
} \label{fig_rgb}
\end{figure*}

Next, we return to pore filling. To gain insight into the elemental composition of the imaged sample area over time, we acquired sequential EELS maps. Each map was recorded over a 4$\times$4~nm$^{2}$ area at methane partial pressures of (1.0 to 1.9)$\times$10$^{-7}$~Torr and across the 170--440~eV energy-loss range. This range was chosen to simultaneously observe the evolution of the B--$K$ edge at 180~eV, the C--$K$ edge at 284~eV, and the N--$K$ edge at 402~eV. The energy scale was recalibrated using the separation of the B-- and N--$K$ edge $\pi^*$ peaks, yielding a dispersion of 0.51~eV/channel. All EELS maps were acquired with a dwell time of 50~ms and at 64$\times$64~pixels, resulting in a recording time of 3.4~min per map. The total duration of each experiment was between 25--35 min.

As shown in Fig.~\ref{fig_rgb}(a), carbon atoms that are progressively incorporated into the lattice over time tend to cluster together. In most cases, these patches appear to form via pore filling as shown above. However, particularly for smaller structures, a direct correspondence between a pore and carbon inclusion cannot always be established, likely because pore formation and healing occur on timescales shorter than the acquisition time of a single map. As shown in Fig.~\ref{fig_rgb}(b), across all five analyzed datasets, the amount of carbon in the first recorded EELS map is significantly lower than in the final frame. However, carbon incorporation into the lattice is generally not constant over time, leading to variations in the absolute amount of carbon present.

%25.April:One important parameter influencing the C growth rate is the arrival of carbon-containing species at the pore edge, described by the impingement rate

%10.Feb:To monitor changes in lattice composition over time, we recorded stacks of EELS maps with a 4 $\times$ 4~nm$^2$ FOV for a total duration of $\sim$30~min, along with ADF images taken between maps for structural information about the lattice. As shown in Fig.~\ref{fig_rgb}, carbon gradually incorporates into the hBN lattice by replacing boron and nitrogen atoms at the edges of the pores, leading to the formation of carbon clusters. During the first $\sim$20~min of irradiation, the carbon signal increases roughly in a linear manner at an average rate of X. This rate is determined by the arrival of carbon-containing species at the pore edge, described by the impingement rate

%25.April\begin{equation}
%J = \frac{10 \times P}{\sqrt{2\pi m k_{\mathrm{B}} T}},
%\end{equation}

%25.April:where $P$ is the pressure, $m$ is the molecular mass (for methane, 16 amu), $k_B$ is the Boltzmann constant, and $T$ is the temperature. The impinging rate is multiplied by 2 for both surfaces of the monolayer (bottom and top), the frame time, and area to determine the number of impinging molecules per frame ($J_{FOV}$). For a gauge pressure of X, $J_{FOV}$ is $\sim$ X. Comparing this value with the average rate of carbon increase suggests that roughly every Xth CH$_4$ molecule hitting the hBN lattice results in a carbon atom being incorporated.

As the carbon concentration in the lattice increases over time, the relative concentrations of boron and nitrogen decrease, as shown in Fig.~\ref{fig_rgb}(b). Similar to carbon incorporation, these changes are not uniform across all analyzed datasets. Nevertheless, a consistent trend emerges in which nitrogen is depleted from the lattice slightly more readily than boron. This observation aligns with previous experiments on the effect under a hydrogen atmosphere, which indicate preferential etching of nitrogen (in contrast to preferential etching of boron observed under an oxygen atmosphere~\cite{JavedInfluenceIrradiation}). However, it is worth noting that the absolute nitrogen concentration is associated with greater uncertainty because its $ K$-edge is less intense than that of boron.

Having established that carbon incorporates into the lattice during the experiment, we turn to the question of spatial confinement. Overview EELS maps covering a 16$\times$16~nm$^2$ field of view were acquired immediately before and after each central 4$\times$4~nm$^2$ irradiation series, providing a direct measure of how far carbon spreads beyond the directly imaged area. In the pre-irradiation maps, carbon covers 0.4$\pm$0.1\% of the imaged lattice area. After irradiation, the amount of carbon increases across all datasets; however, as noted above, the absolute rate of increase varies across datasets. Considering all C atoms in the post-irradiation maps individually, 29$\pm$4\% fall within the irradiated region and 72$\pm$4\% fall outside it, with those outside located, on average, 4.7$\pm$0.5 nm from the irradiated boundary. This diffusion tail reflects the mobility of isolated carbon atoms under the electron beam and sets a practical limit on the precision of single-atom placement with the current approach. However, the picture changes when the analysis is restricted to carbon-dense regions---areas of 0.5$\times$0.5~nm$^2$ containing three or more carbon atoms each. Of such regions, 84$\pm$7\% are located within the irradiated area, and only 16$\pm$7\% are outside. This distinction between single-atom diffusion and confinement of carbon-filled patches is important: the defects responsible for single-photon emission in hBN are thought to involve specific multi-atom carbon configurations~\cite{Li2022UltravioletClusters}, and our results suggest that such configurations form preferentially within the imaged area.

\begin{figure*}[t]
	\centering
	\includegraphics[width=\textwidth]{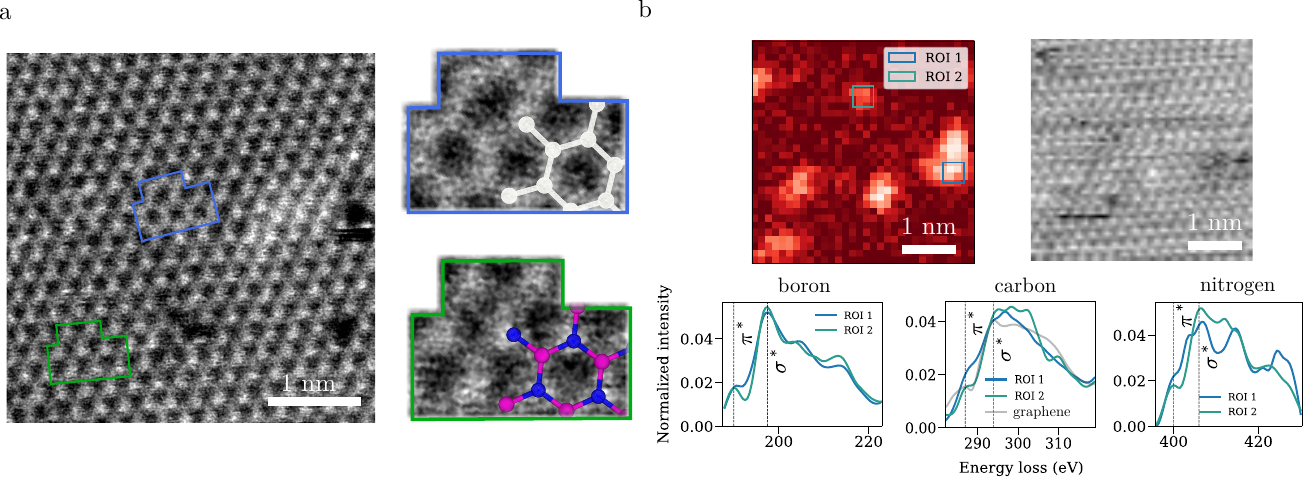}
	\caption{Atomic structure and EELS fine structure of carbon patches formed in hBN. (a) High-resolution MAADF image of the hBN lattice after approximately 2.5~h of continuous electron irradiation at a methane pressure of 1.2$\times$10$^{-7}$~Torr. The blue frame highlights a region containing carbon atoms arranged in a hexagonal pattern; in the inset, white circles indicate carbon atoms. The green frame marks a pristine hBN region, where pink and blue circles respectively represent boron and nitrogen atoms. (b) Elemental map derived from the integrated C--$K$ edge signal, together with representative EELS spectra extracted from selected regions of interest, marked on the elemental map. Spectra are normalized to their integrated area for comparison.} \label{fig_fin}
\end{figure*}

To atomically resolve the carbon patches and characterize their structure, we continuously imaged a 4$\times$4~nm$^2$ FOV under 1.0$\times$10$^{-7}$~Torr methane partial pressure by recording $256\times256$-pixel frames with a dwell time of 16~\textmu s and an electron dose rate of 4.9$\times$10$^{7}$~e$^-/$nm$^2$s. This dose rate was chosen to minimize pore formation during irradiation. In contrast to pore edges, where dynamic structural evolution under the electron beam prevents the acquisition of high signal-to-noise images, this limitation is absent within the basal plane. After approximately 2.5~h of continuous imaging, a high-resolution image of the exposed area was acquired with 1024$\times$1024 pixels. This experiment was performed twice at different positions. As shown in Fig.~\ref{fig_fin}(a), the atomically resolved images reveal patches of carbon atoms, consistent with the EELS-mapping results. Further, the images indicate that the carbon atoms preferentially adopt a hexagonal arrangement within the lattice, analogous to that in graphene.

To connect structural observations to changes in the electronic properties, we analyzed the EELS fine structure of carbon patches identified in the high-resolution imaging experiments. Regions of interest of 0.4$\times$0.4~nm$^2$ were placed over carbon-containing areas of varying sizes, and the fine-structure fits within each region were averaged to reduce noise. The results reveal a clear dependence of the local electronic environment on patch size. In smaller areas, the $\pi^*$ peaks of the B--$K$, C--$K$, and N--$K$ edges are well pronounced (Fig.~\ref{fig_fin}(b), ROI 2), consistent with an $sp^2$-bonded environment analogous to pristine hBN and graphene, respectively, where the $\pi^*$ feature reflects delocalized bonding within the honeycomb lattice. In the central regions of larger patches with a nominal diameter of $\sim$1~nm, however, the $\pi^*$ intensities of both the B--$K$ and C--$K$ edges are suppressed relative to this reference (Fig.~\ref{fig_fin}(b), ROI 1), indicating a disruption of the local bonding environment that cannot be attributed to pristine hBN or graphene-like carbon alone. This modification is consistent with the formation of a carbon-rich phase in which the surrounding boron and nitrogen sublattices are locally perturbed---a structural situation that is expected to generate localized electronic states within the hBN bandgap~\cite{Qiu2024AtomicFunctionalities}. Whether such states give rise to optically active transitions at the single-photon level will require correlated optical measurements, but the observed fine-structure changes confirm that the incorporated carbon meaningfully alters the local electronic structure in a manner that is prerequisite for defect-based quantum emission.

\section*{Conclusion}

We demonstrate a method for localized carbon doping of hexagonal boron nitride facilitated by electron-beam irradiation in a low-pressure methane atmosphere.
We find that increasing the methane partial pressure suppresses pore growth and leads to the formation of boron-terminated triangular pores. Such structures, especially tetravacancies, are even more stable under a H$_2$ atmosphere, suggesting that they form due to preferential etching of nitrogen caused by hydrogen. In contrast, as revealed by time-resolved electron energy-loss spectroscopy maps, carbon dissociated from methane fills vacancies and pores created by boron and nitrogen depletion during electron irradiation. The incorporated carbon forms nanoscale patches that can be clearly attributed to electron-beam–induced doping, as shown by the strong spatial confinement of carbon-doped regions to the areas imaged with the electron beam.
Additionally, we find that the carbon patches preferably adopt a hexagonal arrangement within the lattice and do not exceed $\sim$1~nm in size. Changes in the EELS fine structure further indicate modifications to the local electronic environment within these clusters.

\section*{Methods}

\subsection*{Sample preperation}

Samples were prepared using a chemical delamination method described in Ref.~\cite{Irschik2026AtomicallyVacuum}. Commercial CVD-grown hBN on copper foil (Sigma-Aldrich) was cut into approximately 0.5$\times$0.5~cm$^2$ squares, slightly larger than the transmission electron microscopy (TEM) grids used as the final support. To stabilize the hBN during transfer, the surface was spin-coated with a thin layer of polymethyl methacrylate (PMMA, AR-P 642.04). Spin coating was performed in three stages: 500~rpm during dispensing, 1000~rpm for 15~s, and 2000~rpm for 60~s. The coated copper/hBN/PMMA stacks were then baked on a hot plate at 120$\si{^{\circ}}$C for 10~min to harden the PMMA and improve adhesion to the hBN layer. After baking, the edges and one corner of the foil were trimmed with a surgical scalpel.

The hBN/PMMA layer was separated from the copper foil via electrochemical delamination. A platinum wire acted as the anode, while the copper foil, which was picked up at the scratched corner with inverse tweezers, served as the cathode in a NaOH solution (1~g NaOH in 25~g deionized water). Applying a 4.8~A bias caused hydrogen evolution on the copper surface, which detached the hBN/PMMA stack from the copper substrate. The detached film was gently rinsed in deionized water and scooped out using a Quantifoil TEM gold grid with 1.2~\textmu m radius holes and 1.3~\textmu m center-to-center spacing.

In the final step, the grids were placed on a glass slide and heated to 120$\si{^{\circ}}$C for 12~min to dry the sample and ensure the adhesion of the hBN to the Quantifoil. The PMMA was then removed by immersing the grids in an acetone bath at 50$\si{^{\circ}}$C for 1~h, followed by transferring them to isopropanol for 30~min to remove residual acetone and prevent damage from rapid evaporation. Upon insertion into the CANVAS~\cite{Mangler2022AScale}, the samples were baked at ca. 150$^\circ$C for ca.~12~h in vacuum.

\subsection*{Microscopy and spectroscopy}

All experiments were performed using the aberration-corrected Vienna Nion UltraSTEM 100 operated at 60~kV, with a beam current estimated at 125~pA, as reported in Ref.~\cite{Speckmann2023CombinedMoS2}. The electron probe was formed with a convergence semi-angle of 34~mrad, yielding an effective probe size of $\sim$1.1~\AA. This customized STEM instrument enables gas introduction into the column without compromising atomic-resolution imaging. Using this system, stable gas pressures of up to approximately 2$\times$10$^{-6}$~Torr can be achieved during imaging \cite{Leuthner2019ScanningAtmospheres}. As noted in the main text, although this microscope has a base pressure of $1.5\times 10^{-10}$~Torr, it was about one order of magnitude higher during these experiments due to a minor leak. Under these conditions, we expect the major molecular species in the residual vacuum to be water when the leak valve is closed.

Imaging was performed using either a medium-angle (MAADF) or high-angle annular dark-field (HAADF) detector, with collection semi-angular ranges of 60--200~mrad and 80--300~mrad, respectively. The MAADF image series shown in Fig.~\ref{fig_growth} was recorded over a FOV of 4$\times$4 nm$^2$ with a dwell time of 16 \textmu s and 512$\times$512 pixels. Images were recorded at a partial pressure of 1.1$\times$10$^{-9}$ Torr (leak valve closed), methane partial pressures of 8.0$\times$10$^{-9}$ to 1.9$\times$10$^{-7}$~Torr, and a hydrogen partial pressure of 1.1$\times$10$^{-7}$ Torr, as measured by the objective-area pressure gauge.

Electron energy-loss spectroscopy (EELS) was conducted using a Gatan PEELS~666 spectrometer connected to an Andor iXon~897 electron-multiplying CCD camera. The spectrometer has 512 channels, a dispersion of 0.51~eV/channel, as revealed by the separation of the $\pi^*$ peaks of the B-- and N--$K$ edges, and a collection semi-angle of 35~mrad~\cite{susi_single-atom_2017}. All EELS maps were recorded with a 64$\times$64~pixels and a dwell time of 50~ms, resulting in a recording time of 3.4~min per map. The FOVs were 4$\times$4~nm$^2$ for EELS map series and 16$\times$16 nm$^2$ for the overview maps, resulting in electron doses of 1.0$\times$10$^{-10}$ and 6.2$\times$10$^{-8}$~e$^-/\si{nm^2}$ per map, respectively. The energy window for every map was set to 170--440~eV to simultaneously capture the evolution of the boron ($\sim$188~eV), carbon ($\sim$284~eV), and nitrogen ($\sim$402~eV) $K$ edges. EELS maps were recorded at methane partial pressures of (1.0 to 1.9)$\times$10$^{-7}$~Torr.

\subsection*{Electron ptychography}

For single-sideband (SSB) electron ptychography time-series acquisitions shown in Fig.~\ref{fig_growth}(c), we recorded concurrent HAADF images and convergent-beam electron diffraction patterns for each of the 512$\times$512 scan positions with a pixel dwell time of 10~\textmu s and a maximum scattering angle of 36~mrad on a Dectris ARINA detector binned by 16~\cite{susi_quantifying_2025}. All acquisitions were obtained under a methane partial pressure of 5.5$\times$10$^{-8}$ Torr. The SSB phase reconstructions used the quantitative electron microscopy (quantEM) data analysis toolkit~\cite{varnavides_2025} (version 0.1.8~\cite{quantem}). Residual electron-optical aberrations were corrected post-acquisition by least-squares fitting of their coefficients recursively up to 5th order using six $\mathbf{q}$-vectors in the double-disk overlap regions, with a signal weight of 0.25.

%Barbara8.MaychangeFor single-sideband (SSB) ptychography time-series acquisitions shown in Fig.~\ref{fig_triangle}, we recorded concurrent HAADF images and convergent-beam electron diffraction patterns for each of the 512$\times$512 scan positions with a pixel dwell time of 10~\textmu m and a maximum scattering angle of 36~mrad on a Dectris ARINA detector binned by 16~\cite{susi_quantifying_2025}. The SSB phase reconstructions used the quantitative electron microscopy (quantEM) data analysis toolkit~\cite{varnavides_2025} (version 0.1.8~\cite{quantem}). Residual electron-optical aberrations were corrected post-acquisition by least-squares fitting of their coefficients recursively up to 5th order using six $\mathbf{q}$-vectors in the double-disk overlap regions, with a signal weight of 0.25.

\subsection*{Data processing}

\subsubsection*{ADF images}

To quantify pore-growth rates, MAADF image stacks were analyzed using a semi-automated Python script, further described in Ref.~\cite{JavedInfluenceIrradiation}. For each frame, pores were identified by intensity thresholding, with low-intensity regions corresponding to vacancies, which were segmented and grouped into individual pores. The total pore area per image was obtained by summing the areas of all detected pores within the FOV. Scale calibration was performed using the lattice periodicity extracted from the fast Fourier transform (FFT) of the images, providing an accurate nanometer-per-pixel conversion. This calibration was then used to convert pore sizes from pixel units to physical areas.

For the filtered HAADF images in Fig.~\ref{fig_growth}(c), after equalizing the image histograms to enchance contrast, we applied a double-Gaussian filter (with parameters $\sigma_1$ = 9~px, $\sigma_2$ = 6~px, $w$ = 0.3), followed by a Butterworth high-pass filter to suppress low-frequency intensity variation presumably due to varying beam current (order 1, with a spatial frequency of 0.1/px).

\subsubsection*{EELS maps}

EELS maps were processed using the HyperSpy/eXSpy packages (versions 2.4.0 and 0.3.2, respectively)~\cite{Hyperspy,Exspy}. Single-pixel intensity spikes attributed to cosmic-ray hits were removed before spatial binning by a factor of 2, reducing the map dimensions from 64$\times$64 to 32$\times$32 pixels. The spectra were then convolved with a 2~eV Gaussian to reduce noise in the fine-structure signals, and the energy axis calibrated by aligning the measured $\pi^*$ peak positions of the B-- and N--$K$ edges with a reference spectrum from EELS.info~\cite{EELS.info}, yielding a dispersion of 0.51 eV/channel (maximum allowed by our spectrometer). For each pixel, we first fitted the core-loss edges without a PowerLaw background using the `dft' edge shape, with fitting windows of 187.5--223~eV, 282--319~eV, and 395.7--430.2~eV for the B--$K$, C--$K$, and N--$K$ edges, respectively. We then enabled the background and refitted; this two-step fitting procedure was found to improve numerical robustness. Fine structures for each edge were then iteratively fitted using the \texttt{multifit(kind=\textsf{'}smart\textsf{'})} built-in functionality of HyperSpy/eXSpy, and the intensities of each edge were automatically corrected by the oscillator strengths for each core-level transition to obtain the elemental composition. Finally, negative edge intensities were constrained to positive values, and the model was refitted.

To convert fitted edge intensities into atom counts, the first map of each series served as a pristine hBN reference. The expected number of B and N atoms within the field of view was calculated geometrically from the hBN lattice constant ($a$ = 0.2504 nm), yielding approximately 311 atoms per sublattice in the 4$\times$4 nm$^2$ maps and approximately 4715 atoms per sublattice in the 16$\times$16 nm$^2$ overview maps. The per-atom intensity for B and N was then determined by selecting a 10$\times$10-pixel ROI in a visually defect-free, carbon-free area of the first frame and dividing the total fitted edge intensity within that ROI by the corresponding expected atom count. The uncertainty was taken as the standard error of the mean (SEM) of the per-pixel intensities within the ROI. Since carbon is not present in each pixel, and because the C--$K$ edge ionization cross-section lies between those of B and N, the carbon per-atom intensity was taken as the average of the B and N values, with the uncertainty propagated as the average of their respective SEMs.

%21.MayThe EELS maps were processed using an interactive procedure with the Hyperspy/Exspy packages~\cite{Hyperspy,Exspy} (versions 2.4.0 and 0.3.2). The energy scale was recalibrated based on the B-- and N--K edge $\pi^*$ peak separation, resulting in a dispersion of 0.51~eV/channel (maximum allowed by our spectrometer). After removing single-pixel intensity spikes presumably due to cosmic rays, the spectra were convolved with a Gaussian of width 2~eV to reduce the noise of the fine-structure signals. For each pixel, we then first fitted the core-loss edges without a PowerLaw background using the `dft' edge shape, and then enabled the background and refitted. Fine-structures for each edge were finally iteratively fitted using the `multifit(kind=\textsf{'}smart\textsf{'})` using the built-in functionality of HyperSpy/Exspy, and the intensities of each edge automatically corrected by the oscillor strengths for each core-level transition to obtain the elemental composition for each element. %To reduce noise and isolate relevant features, masking procedures were applied to the integrated maps. This included threshold-based segmentation to identify element-specific signal regions, followed by morphological operations to remove isolated pixels and small features below the spatial resolution limit. The resulting masked maps were used for all quantitative analysis of the concentration and spatial distributions of B, C, and N. 

\section*{Acknowledgements}

%Insert acknowledgements 

ML acknowledges funding from the European Union’s Horizon Europe research and innovation programme under the Marie Skłodowska-Curie grant agreement No. 101210084.
This research was funded in part by the Austrian Science Fund (FWF) [10.55776/COE5]. For open-access purposes, the author has applied a CC-BY public copyright license to any author-accepted manuscript version arising from this submission.

\bibliography{apssamp}% Produces the bibliography via BibTeX.

\clearpage

\twocolumngrid
\onecolumngrid

\begin{center}
    \large {Supplemental Material}\\[1em]
    \large \textbf{Electron-beam-induced carbon doping of hexagonal boron nitride}\\
\end{center}
\clearpage
\renewcommand{\thefigure}{S\arabic{figure}}
\setcounter{figure}{0}  % Restart figure numbering from S1

\begin{figure}[htbp]
    \centering
    \includegraphics[width=1\textwidth]{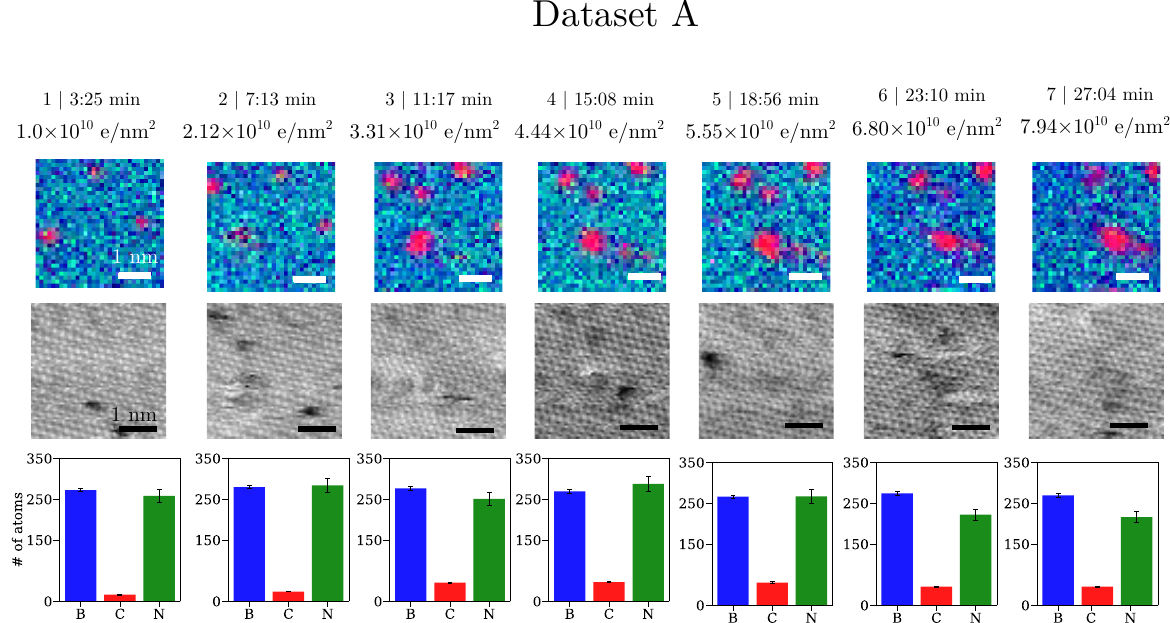}
    \caption{Temporal evolution of carbon incorporation and concurrent boron and nitrogen depletion in hBN under electron-beam irradiation in a methane atmosphere at (1.0 to 1.9)$\times 10^{-7}$~Torr for Dataset A. (a) The first row shows elemental maps derived from the integrated B--, C--, and N--$K$ edge signals respectively represented in blue, red, and green. The second row presents the corresponding HAADF images acquired simultaneously with the elemental maps. The third row shows the quantified numbers of B, C, and N atoms extracted from each frame, illustrating the progressive incorporation of carbon together with the corresponding depletion of boron and nitrogen over time.} 
    \label{fig: s1}
\end{figure}

\begin{figure}[htbp]
    \centering
    \includegraphics[width=1\textwidth]{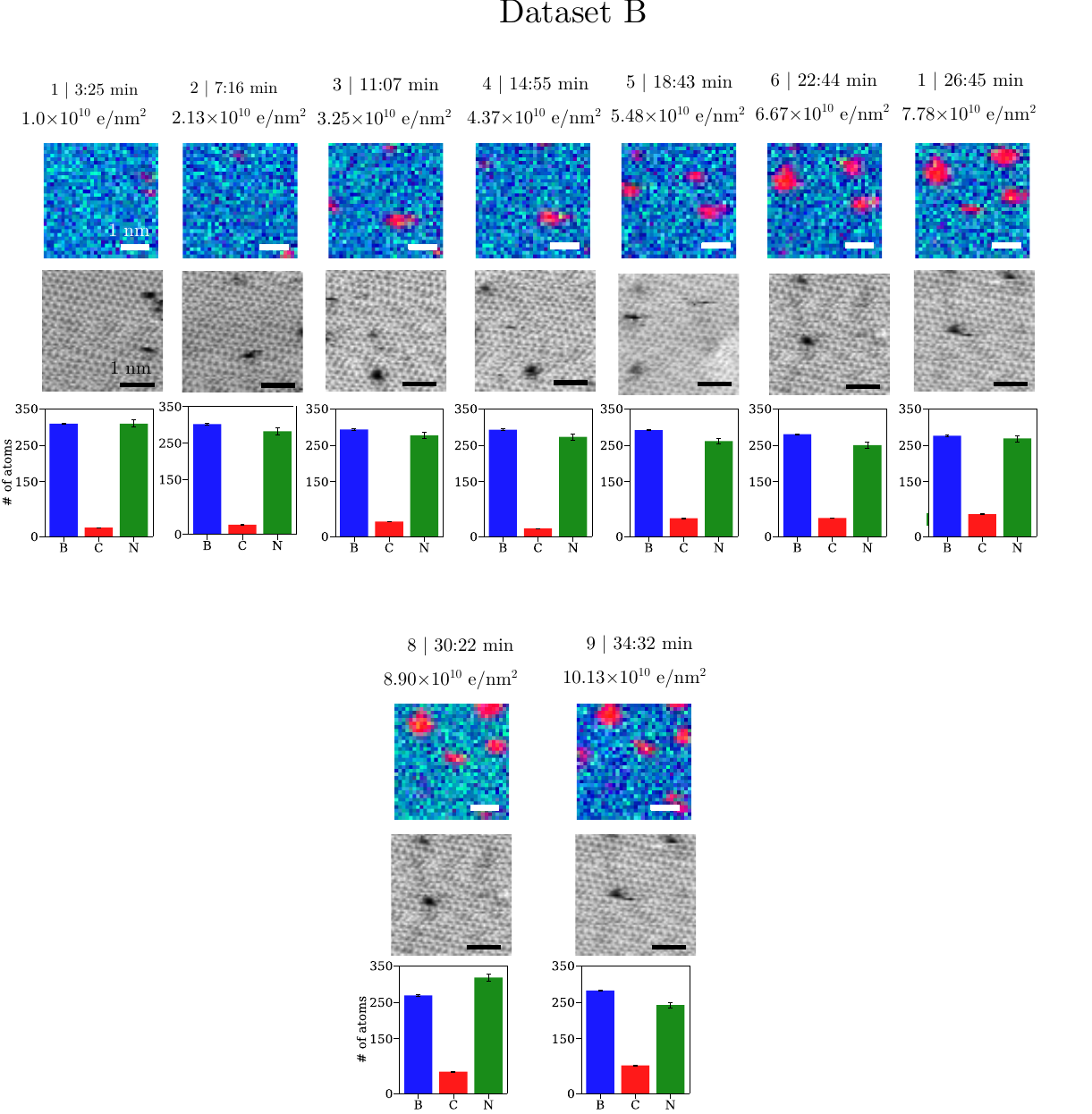}
    \caption{Temporal evolution of carbon incorporation and concurrent boron and nitrogen depletion in hBN under electron-beam irradiation in a methane atmosphere at (1.0 to 1.9)$\times 10^{-7}$~Torr for Dataset B. (a) The first row shows elemental maps derived from the integrated B--, C--, and N--$K$ edge signals respectively represented in blue, red, and green. The second row presents the corresponding HAADF images acquired simultaneously with the elemental maps. The third row shows the quantified numbers of B, C, and N atoms extracted from each frame, illustrating the progressive incorporation of carbon together with the corresponding depletion of boron and nitrogen over time.} 
    \label{fig: s2}
\end{figure}

\clearpage

\begin{figure}[htbp]
    \centering
    \includegraphics[width=1\textwidth]{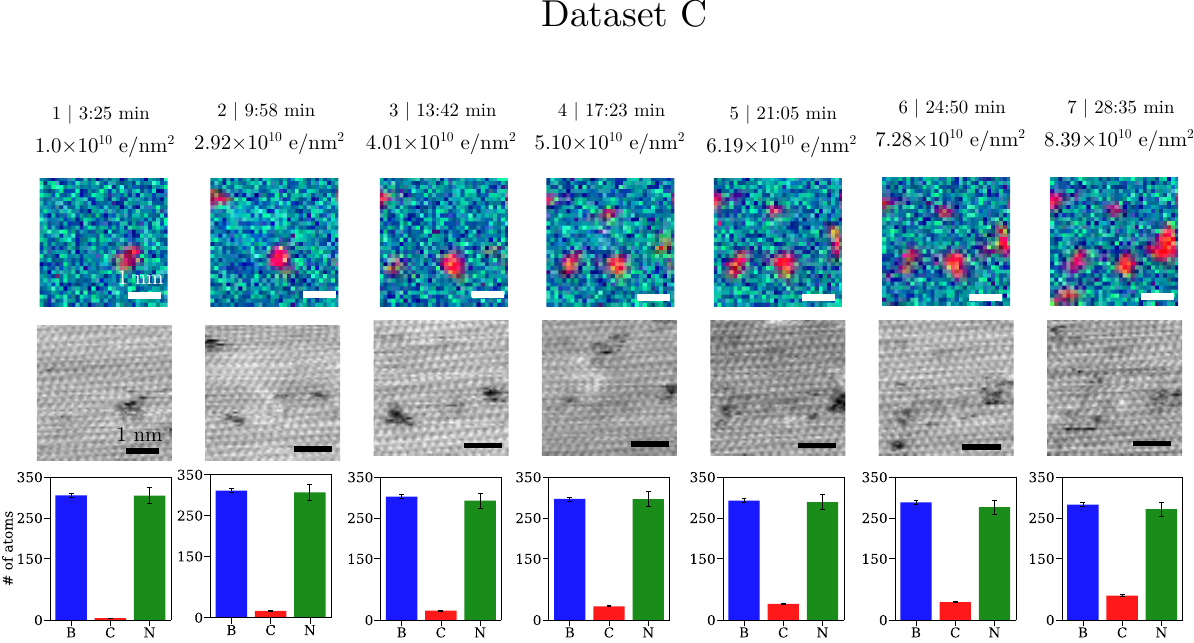}
    \caption{Temporal evolution of carbon incorporation and concurrent boron and nitrogen depletion in hBN under electron-beam irradiation in a methane atmosphere at (1.0 to 1.9)$\times 10^{-7}$~Torr for Dataset C. (a) The first row shows elemental maps derived from the integrated B--, C--, and N--$K$ edge signals respectively represented in blue, red, and green. The second row presents the corresponding HAADF images acquired simultaneously with the elemental maps. The third row shows the quantified numbers of B, C, and N atoms extracted from each frame, illustrating the progressive incorporation of carbon together with the corresponding depletion of boron and nitrogen over time.} 
    \label{fig: s3}
\end{figure}

\begin{figure}[htbp]
    \centering
    \includegraphics[width=1\textwidth]{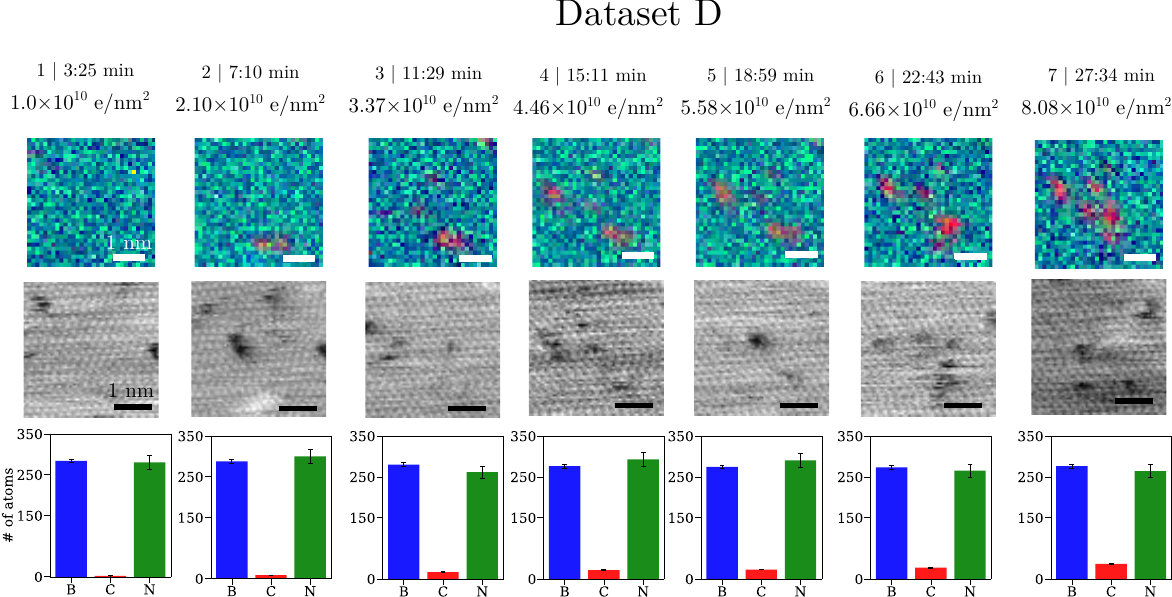}
    \caption{Temporal evolution of carbon incorporation and concurrent boron and nitrogen depletion in hBN under electron-beam irradiation in a methane atmosphere at (1.0 to 1.9)$\times 10^{-7}$~Torr for Dataset D. (a) The first row shows elemental maps derived from the integrated B--, C--, and N--$K$ edge signals respectively represented in blue, red, and green. The second row presents the corresponding HAADF images acquired simultaneously with the elemental maps. The third row shows the quantified numbers of B, C, and N atoms extracted from each frame, illustrating the progressive incorporation of carbon together with the corresponding depletion of boron and nitrogen over time.} 
    \label{fig: s4}
\end{figure}

\begin{figure}[htbp]
    \centering
    \includegraphics[width=1\textwidth]{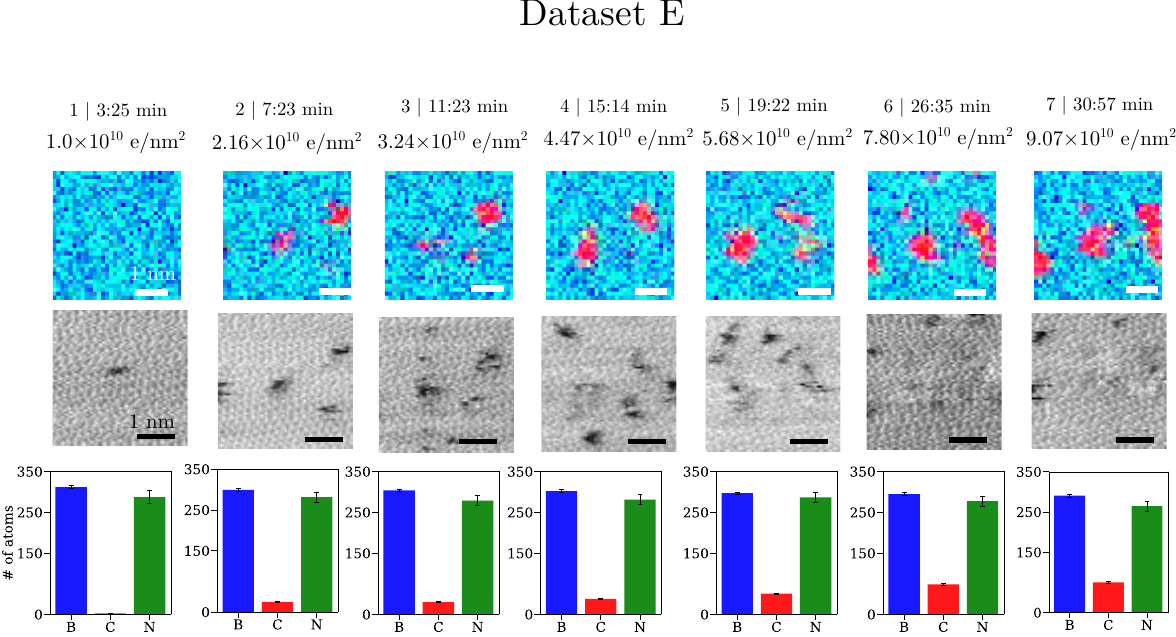}
    \caption{Temporal evolution of carbon incorporation and concurrent boron and nitrogen depletion in hBN under electron-beam irradiation in a methane atmosphere at (1.0 to 1.9)$\times 10^{-7}$~Torr for Dataset E. (a) The first row shows elemental maps derived from the integrated B--, C--, and N--$K$ edge signals respectively represented in blue, red, and green. The second row presents the corresponding HAADF images acquired simultaneously with the elemental maps. The third row shows the quantified numbers of B, C, and N atoms extracted from each frame, illustrating the progressive incorporation of carbon together with the corresponding depletion of boron and nitrogen over time.} 
    \label{fig: s5}
\end{figure}

\clearpage

\begin{figure}[htbp]
    \centering
    \includegraphics[width=0.6\textwidth]{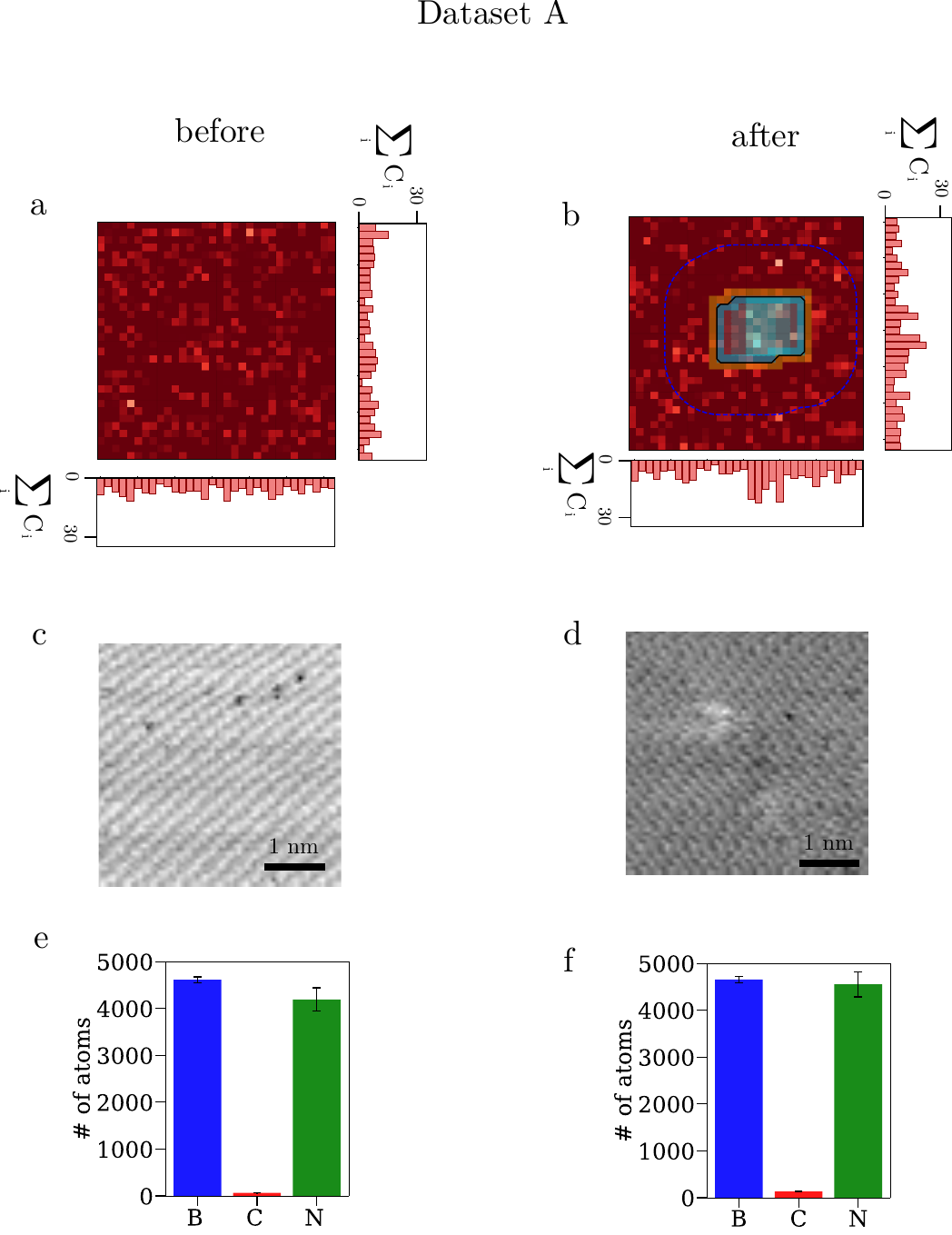}
    \caption{Overview EELS maps for Dataset A. (a) Heatmap showing the carbon abundance across a 16$\times$16 nm$^2$ field of view (FOV), acquired before the 4$\times$4 nm$^2$ FOV EELS map series shown in Fig.~\ref{fig: s1}. The blue-highlighted region enclosed by the black outline indicates the area exposed to the electron beam, while the orange region represents the uncertainty in the electron-beam position. The blue dashed line indicates the average distance between carbon atoms located outside the electron-beam-irradiated region and the irradiated region itself. The histograms below and to the right of the heatmap show the respective summed carbon-atom counts for the columns and rows of the map. (b) Heatmap showing the carbon abundance across a 16$\times$16 nm$^2$ FOV, acquired after the 4$\times$4 nm$^2$ FOV EELS map series shown in Fig.~\ref{fig: s1}. (c)--(d) Concurrent HAADF images corresponding to the overview maps shown in (a) and (b). (e)--(f) Amounts of boron, carbon, and nitrogen detected in the EELS maps acquired before and after the central map series.} 
    \label{fig: s6}
\end{figure}

\begin{figure}[htbp]
    \centering
    \includegraphics[width=0.6\textwidth]{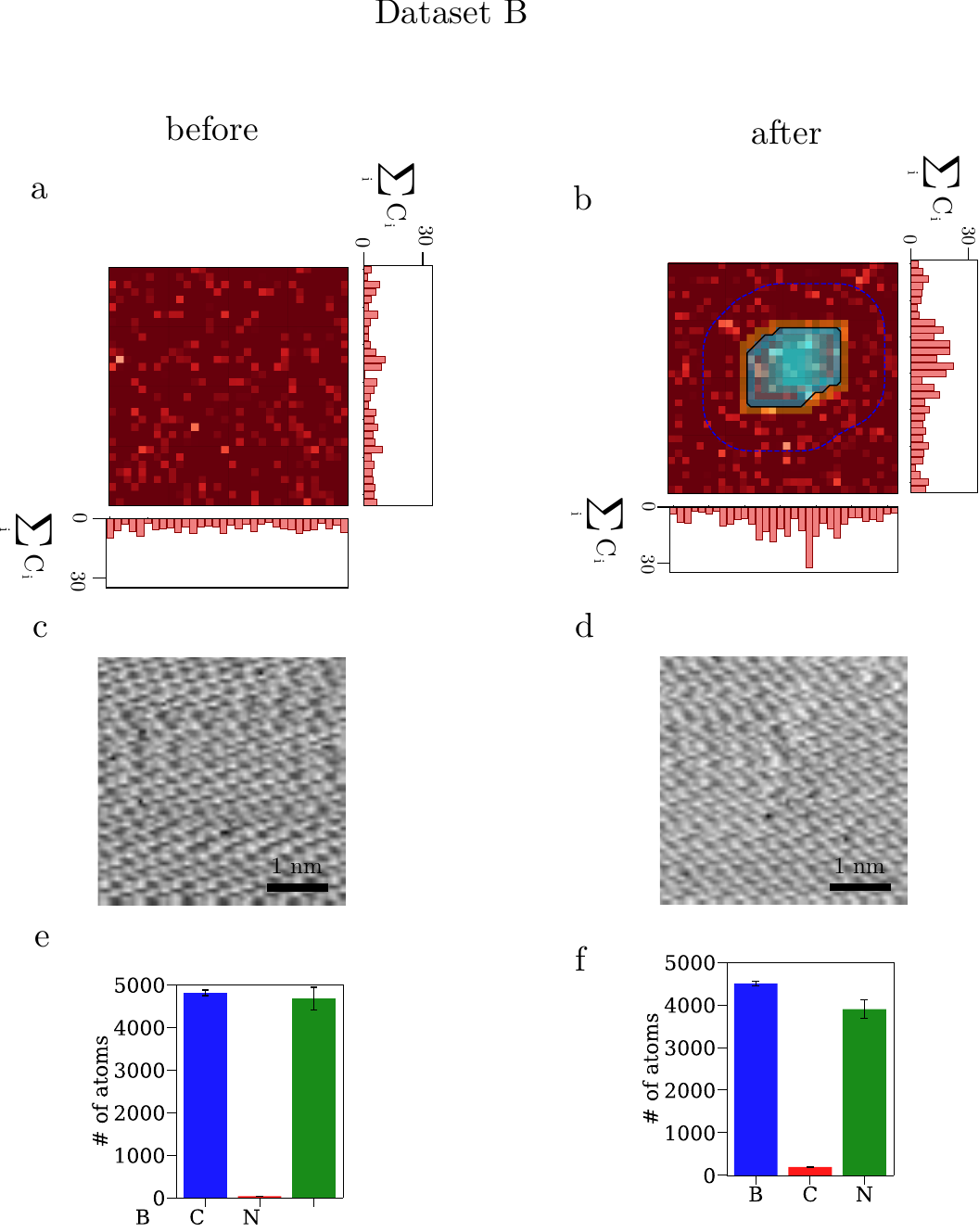}
    \caption{Overview EELS maps for Dataset B. (a) Heatmap showing the carbon abundance across a 16$\times$16 nm$^2$ field of view (FOV), acquired before the 4$\times$4 nm$^2$ FOV EELS map series shown in Fig.~\ref{fig: s2}. The blue-highlighted region enclosed by the black outline indicates the area exposed to the electron beam, while the orange region represents the uncertainty in the electron-beam position. The blue dashed line indicates the average distance between carbon atoms located outside the electron-beam-irradiated region and the irradiated region itself. The histograms below and to the right of the heatmap show the respective summed carbon atom counts for the columns and rows of the map. (b) Heatmap showing the carbon abundance across a 16$\times$16 nm$^2$ FOV, acquired after the 4$\times$4 nm$^2$ FOV EELS map series shown in Fig.~\ref{fig: s2}. (c)--(d) Concurrent HAADF images corresponding to the overview maps shown in (a) and (b). (e)--(f) Amounts of boron, carbon, and nitrogen detected in the EELS maps acquired before and after the central map series.} 
    \label{fig: s7}
\end{figure}

\begin{figure}[htbp]
    \centering
    \includegraphics[width=0.6\textwidth]{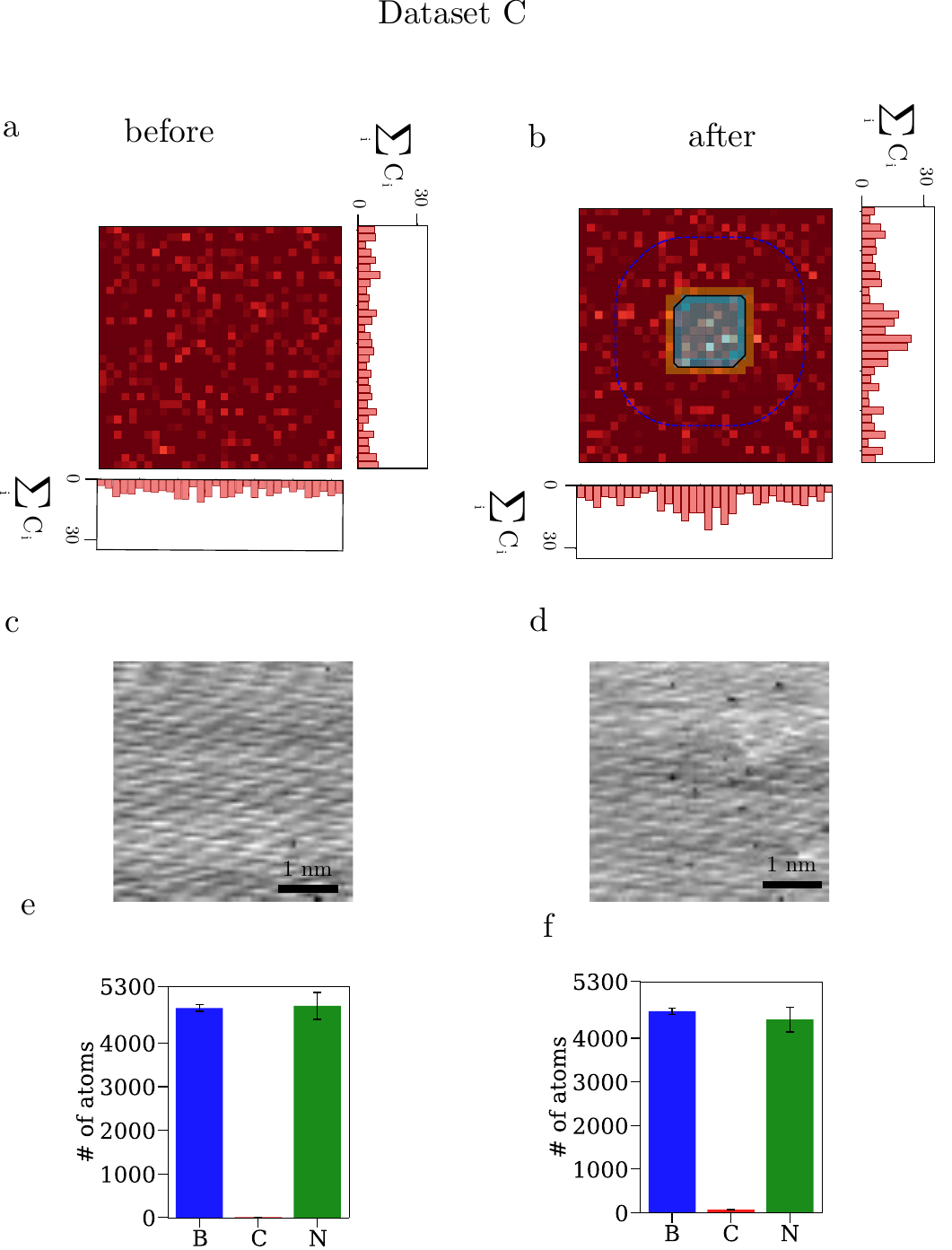}
    \caption{Overview EELS maps for Dataset C. (a) Heatmap showing the carbon abundance across a 16$\times$16 nm$^2$ field of view (FOV), acquired before the 4$\times$4 nm$^2$ FOV EELS map series shown in Fig.~\ref{fig: s3}. The blue-highlighted region enclosed by the black outline indicates the area exposed to the electron beam, while the orange region represents the uncertainty in the electron-beam position. The blue dashed line indicates the average distance between carbon atoms located outside the electron-beam-irradiated region and the irradiated region itself. The histograms below and to the right of the heatmap show the respective summed carbon atom counts for the columns and rows of the map. (b) Heatmap showing the carbon abundance across a 16$\times$16 nm$^2$ FOV, acquired after the 4$\times$4 nm$^2$ FOV EELS map series shown in Fig.~\ref{fig: s3}. (c)--(d) Concurrent HAADF images corresponding to the overview maps shown in (a) and (b). (e)--(f) Amounts of boron, carbon, and nitrogen detected in the EELS maps acquired before and after the central map series.} 
    \label{fig: s8}
\end{figure}

\begin{figure}[htbp]
    \centering
    \includegraphics[width=0.7\textwidth]{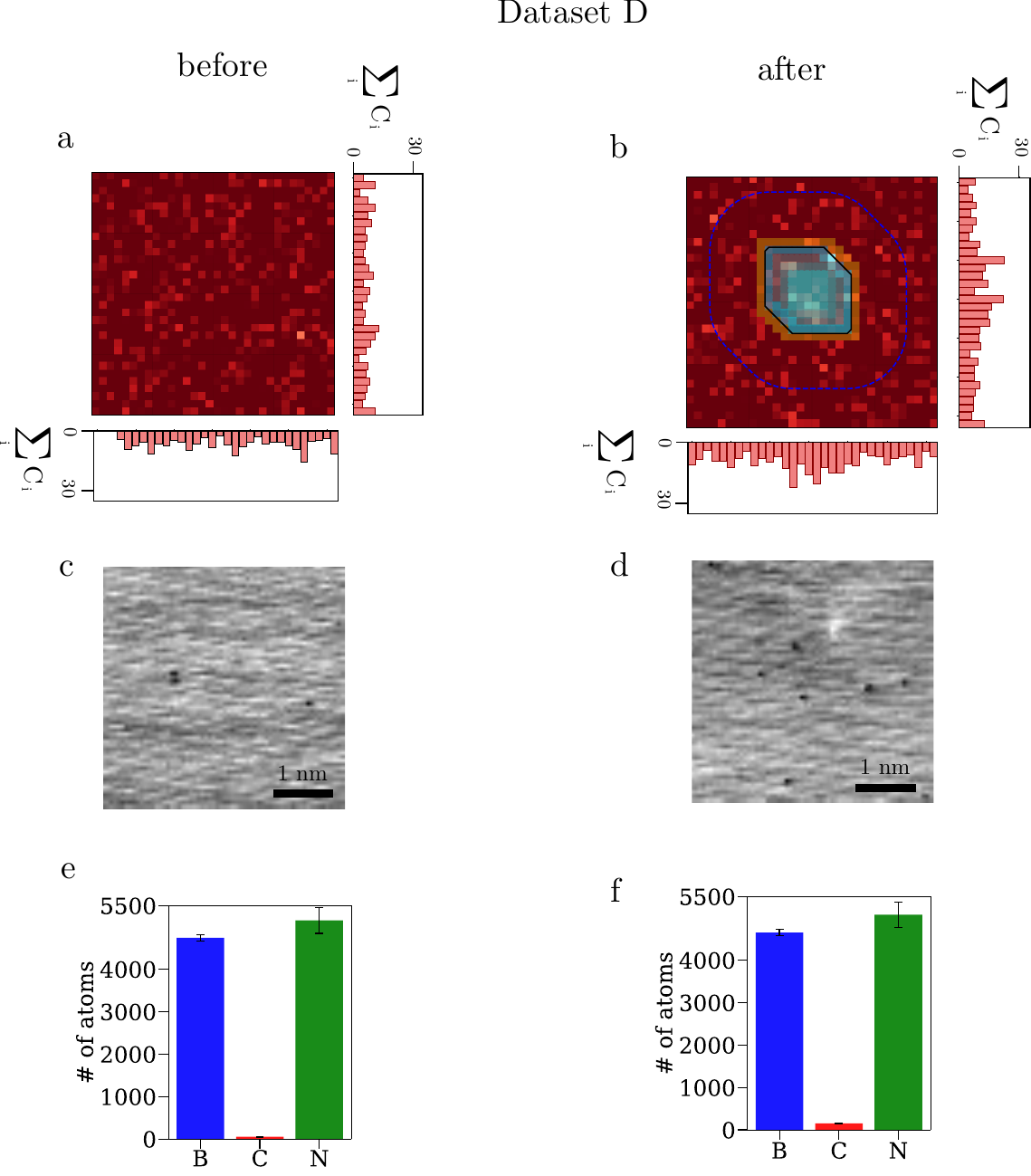}
    \caption{Overview EELS maps for Dataset D. (a) Heatmap showing the carbon abundance across a 16$\times$16 nm$^2$ field of view (FOV), acquired before the 4$\times$4 nm$^2$ FOV EELS map series shown in Fig.~\ref{fig: s4}. The blue-highlighted region enclosed by the black outline indicates the area exposed to the electron beam, while the orange region represents the uncertainty in the electron-beam position. The blue dashed line indicates the average distance between carbon atoms located outside the electron-beam-irradiated region and the irradiated region itself. The histograms below and to the right of the heatmap show the respective summed carbon atom counts for the columns and rows of the map. (b) Heatmap showing the carbon abundance across a 16$\times$16 nm$^2$ FOV, acquired after the 4$\times$4 nm$^2$ FOV EELS map series shown in Fig.~\ref{fig: s4}. (c)--(d) Concurrent HAADF images corresponding to the overview maps shown in (a) and (b). (e)--(f) Amounts of boron, carbon, and nitrogen detected in the EELS maps acquired before and after the central map series.} 
    \label{fig: s9}
\end{figure}

\end{document}